\documentclass[11pt,letterpaper]{article}

\usepackage[margin=1in]{geometry}
\usepackage[T1]{fontenc}
\usepackage{lmodern,microtype}
\usepackage{amsmath,amssymb,amsthm,mathtools}
\usepackage{graphicx,booktabs,array,longtable,float}
\usepackage[font=small,labelfont=bf]{caption}
\usepackage[round,authoryear]{natbib}
\usepackage{xurl}
\usepackage{tikz}
\usetikzlibrary{arrows,arrows.meta,positioning}
\usepackage[colorlinks=true,allcolors=blue!45!black,bookmarksnumbered=true]{hyperref}
\newtheorem{Pro}{Proposition}
\title{CASPER: Discovering Spatially Resolved\\Causal Cell-Cell Communications}
\author{Yang Ni\\[3pt]
  \small Department of Statistics and Data Science\\
  \small The University of Texas at Austin, Austin, TX, U.S.A.\\
  \small \href{mailto:yang.ni@austin.utexas.edu}{\texttt{yang.ni@austin.utexas.edu}}}
\date{}
\hypersetup{pdftitle={CASPER: Discovering Spatially Resolved Causal Cell-Cell Communications},pdfauthor={Yang Ni}}

\begin{document}
\maketitle

\begin{abstract}
Spatial transcriptomics can reveal where ligands and receptors are expressed together. A harder question is whether exposure to a ligand changes pathway activity in nearby cells, since the same tissue environment can influence both ligand expression and pathway activity. We introduce CASPER, a Bayesian structural equation model that adjusts for receptor expression and represents the shared tissue environment with a Gaussian Markov random field. We show that the ligand-exposure effect is identifiable when the spatial covariance is known and distinguishable from independent noise, the coefficients are constant, and the structural errors are independent and homoskedastic. In simulations, CASPER reduced the bias seen in ordinary regression and in a regression that accounted for spatial dependence only in the outcome. We compared CASPER with CellPhoneDB in four public Xenium datasets. In ovarian cancer, CellPhoneDB found evidence of ligand--receptor expression enrichment for both IL6 and VEGFA. CASPER found greater support for a positive effect of fibroblast IL6 on the inflammatory-response score of ciliated epithelial cells than for an effect of fibroblast VEGFA on the angiogenesis score of endothelial cells.

\smallskip
\noindent\textbf{Keywords.} Causal discovery; Markov random fields; Spatial confounding; Spatial transcriptomics; Structural equation models.
\end{abstract}

\section{Introduction}

\begin{figure}[htbp]
    \centering
\includegraphics[width=\linewidth]{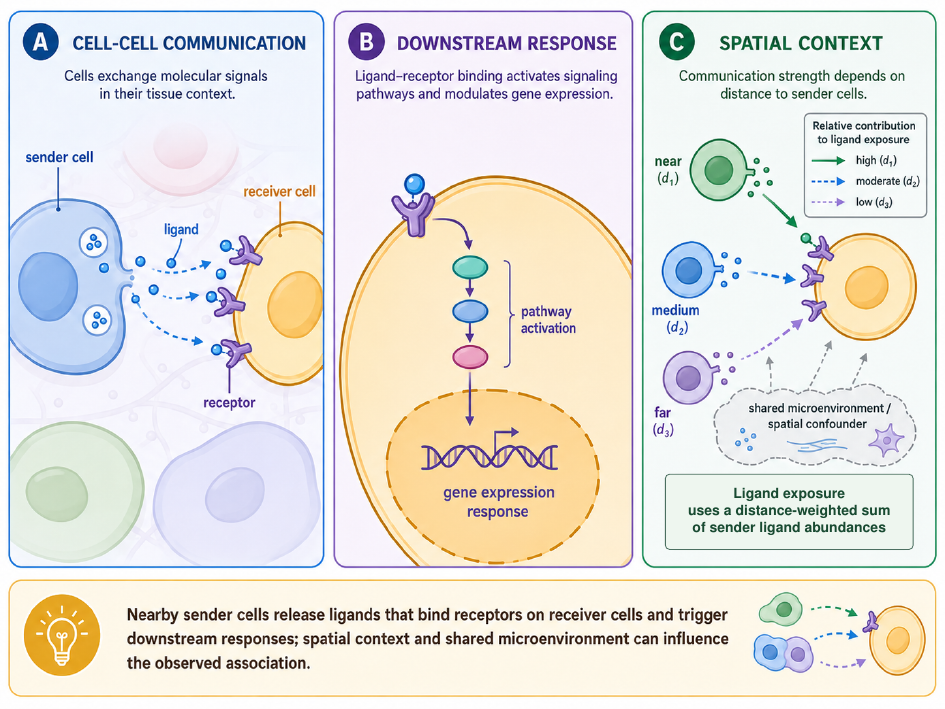}
    \caption{A sender cell releases a ligand that can bind a receptor on a nearby cell (A) and trigger a change in downstream gene expression (B). Nearby senders contribute more to a receiver's exposure than distant senders do, and a shared tissue environment can influence both exposure and response (C). Generative AI was used in producing this illustration.}
    \label{fig:illu}
\end{figure}

Cell-cell communication (CCC) is fundamental to tissue development, immune defense, and disease progression. A ligand released by a sender cell can bind a receptor on a nearby cell and activate a downstream signaling pathway (Figure~\ref{fig:illu}A and B). Spatial transcriptomics measures the expression of the ligand, receptor, and downstream genes in the same tissue. The question is whether changing ligand exposure would change pathway activity in the receiver cell.

Single-cell RNA sequencing and spatial transcriptomics have given rise to tools that prioritize candidate interactions using ligand--receptor co-expression, enrichment, spatial proximity, or target-gene expression \citep{efremova2020cellphonedb,browaeys2020nichenet,jin2021inference,hou2020predicting,cabello2020singlecellsignalr,wang2019italk,dries2021giotto,palla2022squidpy}. A shared tissue environment can produce the same patterns. For example, an inflammatory niche may induce cytokine expression in sender cells and independently activate an inflammatory program in nearby receiver cells (Figure~\ref{fig:illu}C). Methods for inferring directional communication have also been proposed \citep{almet2024inferring,dupuis2025causalccc}. Our interest is in estimating the effect of ligand exposure when an unmeasured feature of the tissue can influence both exposure and response.

We propose \texttt{CASPER}, a Bayesian structural equation model (SEM) for this problem. For a specified ligand, receptor, and pathway, CASPER relates pathway activity to ligand exposure from neighboring senders while adjusting for receptor expression. A shared Gaussian Markov random field represents the influence of the local tissue environment on all three measurements.

The ligand--receptor mechanism suggests a direction of effect, from ligand exposure to pathway activity in the receiver cell. The difficulty is estimating the size of that effect when the same local environment influences both measurements. Work on spatial confounding has shown that the spatial scale and covariance of an unmeasured process can affect a regression estimate \citep{reich2006effects,hodges2010adding,paciorek2010importance,hanks2015restricted}.

CASPER uses the difference between the spatial covariance of the shared field and the independent variation in each measurement to identify the ligand-exposure effect. Irregular cell locations can provide this information through differences in marginal variances. Correlation between cells can also be informative when the marginal variances are constant. We examine these two sources of information in simulations and compare CASPER with CellPhoneDB across four tissue types. Two ovarian examples illustrate how the evidence for a pathway response can differ even when both ligand--receptor pairs show expression enrichment.

\section{Method}\label{sec:method}

For a given sender and receiver cell type, we use prior biological knowledge to specify a sender ligand, a receiver receptor, and a downstream pathway. For example, if fibroblasts are the senders and tumor cells are the receivers, one possible triplet is
\begin{align*}
\underbrace{IL6}_{\mathclap{\substack{\text{fibroblast} \\ \text{ligand}}}} - \overbrace{\underbrace{IL6R}_{\text{receptor}} - \underbrace{\text{inflammatory response}}_{\text{pathway}}}^{\text{tumor}}
\end{align*}
Here the hypothesis is that exposure to fibroblast IL6 changes inflammatory pathway activity in tumor cells expressing IL6R. For receiver $i=1,\ldots,n$, let $X_i$ denote its ligand exposure, $Z_i$ its receptor expression, and $Y_i$ its pathway activity score. We center and scale each variable across receivers before fitting the model.

\paragraph{Ligand exposure via spatial kernel weighting.}
Let $L_k$ be the expression of the ligand in sender cell $k=1,\dots,m$, and let $d_{ik}$ be the Euclidean distance between receiver cell $i$ and sender cell $k$. Define a nonnegative spatial kernel $K(\cdot;h)$ with bandwidth $h>0$. The kernel-weighted exposure of receiver $i$ to the ligand is
\[
E_i \;=\; \sum_{k=1}^m K(d_{ik};h)L_k.
\]

In the applications, $L_k$ is normalized ligand abundance on the linear scale. We use a Gaussian kernel restricted to nearby sender cells and center and scale $\log(1+E_i)$ to obtain $X_i$. Section~\ref{sec:real} gives the neighborhood and bandwidth rules.

\paragraph{Pathway activity score.}
We summarize pathway activity by the mean log-normalized expression of its measured genes. Other possible scores include AUCell \citep{aibar2017scenic} and UCell \citep{andreatta2021ucell}.

\begin{figure}[htbp]
    \centering
    \resizebox{0.9\textwidth}{!}{
    \begin{tikzpicture}[->,>=stealth',node distance=2cm, thick,
        cell/.style={rectangle, draw, rounded corners, align=center, minimum height=1cm, minimum width=2cm}]

        \node[cell] (Y) {$Y_i$ \\ (Pathway activity)};
        \node[cell, right=3cm of Y] (Z) {$Z_i$ \\ (Receptor expression)};

        \node[cell, below=1.5cm of Y] (U) {$U_i$ \\ (Latent neighborhood conditions)};
        \node[cell, left=3cm of Y] (X) {$X_i$ \\ (Ligand exposure)};

        \draw[->] (X) -- (Y);
        \draw[->] (Z) -- (Y);

        \draw[->] (U) -- (X);
        \draw[->] (U) -- (Z);
        \draw[->] (U) -- (Y);

    \end{tikzpicture}
    }
\caption{Structural model for a receiver cell. Ligand exposure ($X_i$) and receptor expression ($Z_i$) affect pathway activity ($Y_i$). The shared spatial field ($U_i$) influences all three measurements.}

    \label{fig:sem-diagram}
\end{figure}
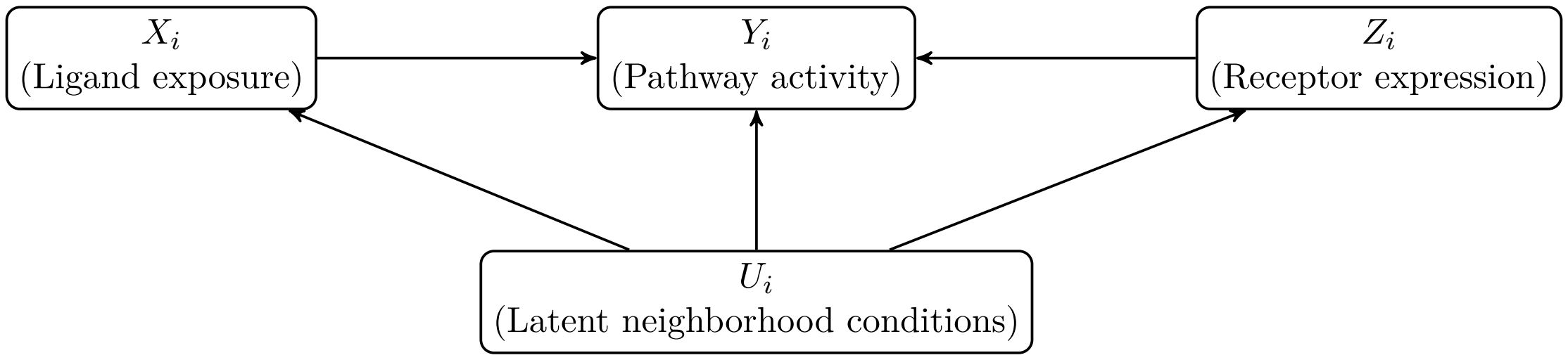

\paragraph{The proposed SEM.}
Figure~\ref{fig:sem-diagram} shows how the proposed model links ligand exposure, receptor expression, and pathway activity in a receiver cell. Let $U_i$ denote unmeasured conditions in the neighborhood of receiver $i$. The structural equations are
\begin{align}\label{eq:sem1}
\begin{aligned}
X_i &=  U_i \beta_{UX} + \epsilon_{X_i}, \\
    Z_i &=  U_i \beta_{UZ} + \epsilon_{Z_i},\\
    Y_i &= X_i \beta_{XY} + Z_i \beta_{ZY} + U_i \beta_{UY} + \epsilon_{Y_i}.
\end{aligned}
\end{align}
The first two equations describe how ligand exposure and receptor expression vary across receivers with local tissue conditions. Exposure is calculated from ligand expression in nearby senders, and receptor expression is measured in the receiver itself. The third equation relates pathway activity to ligand exposure and receptor expression through $\beta_{XY}$ and $\beta_{ZY}$, respectively. It also allows the same local conditions to influence pathway activity directly through $\beta_{UY}$.

The errors are mutually independent across cells and equations and independent of the latent field. For each receiver, $\epsilon_{X_i}\sim\mathcal{N}(0,\sigma_X^2)$, $\epsilon_{Z_i}\sim\mathcal{N}(0,\sigma_Z^2)$, and $\epsilon_{Y_i}\sim\mathcal{N}(0,\sigma_Y^2)$.

Let $Y_i(x,z)$ denote the pathway score that receiver $i$ would have if its exposure and receptor expression were set to $x$ and $z$, with the tissue geometry and latent background held fixed. Under Equation~\eqref{eq:sem1}, the expected change from increasing exposure by $\Delta$ at fixed $z$ is $\mathbb{E}\{Y_i(x+\Delta,z)-Y_i(x,z)\}=\beta_{XY}\Delta$. To relate this contrast to a change in sender expression, write $X_i=g_i(\mathbf L)$, where $\mathbf L$ collects the ligand abundances of the senders. Changing $\mathbf L$ to $\mathbf L'$ can alter the exposure of several receivers at once. If this change acts on receiver $i$ through its exposure summary alone, then at fixed receptor abundance and background conditions the expected change in its pathway score is $\beta_{XY}\{g_i(\mathbf L')-g_i(\mathbf L)\}$.

Let $\theta=(\beta_{UX},\beta_{UZ},\beta_{XY}, \beta_{ZY}, \beta_{UY},\sigma_X^2,\sigma_Z^2,\sigma_Y^2)$ denote all the model parameters.

\paragraph{Modeling spatial confounding.}
Unmeasured neighborhood conditions are represented by a receiver-indexed latent field. Let $\mathbf A$ be the symmetrized, Gaussian-weighted adjacency matrix of the $k$-nearest-neighbor graph on the receivers ($k=8$), and let $\mathbf D$ be the diagonal matrix of weighted degrees. The edge bandwidth is the median positive distance from a receiver to one of its $k$ nearest receiver neighbors. Adding $\kappa=0.01$ to each diagonal entry of the graph Laplacian $\mathbf D-\mathbf A$ gives a positive definite precision matrix $\mathbf Q=\mathbf D-\mathbf A+\kappa\mathbf I$. This defines a proper Gaussian Markov random field (GMRF) \citep{rue2005gaussian} with density
\[
    p(\mathbf{U}) \propto \exp\{-\mathbf{U}^\top \mathbf{Q} \mathbf{U}/2\},
\]
so $\mathbf U\sim\mathcal N(\mathbf 0,\mathbf{C})$ with $\mathbf{C}=\mathbf{Q}^{-1}$. The exposure kernel describes which senders contribute ligand to each receiver. The receiver graph describes how background conditions covary across receivers. The physical range of each therefore depends on the distances between the cells it connects.

The covariance scale is fixed because multiplying $\mathbf{U}$ and dividing its three loadings by the same constant leaves their products unchanged.

\paragraph{Identifiability.}
The known spatial covariance allows the model to distinguish variation shared across cells from variation due to independent errors. The following proposition makes this distinction precise.
\begin{Pro}[Causal Identifiability]\label{prop1}
Suppose $\mathbf{C}$ is known and positive definite, $\mathbf{C}$ and $\mathbf{I}$ are linearly independent, $\beta_{UX}>0$, and $\sigma_X^2,\sigma_Z^2,\sigma_Y^2>0$, with mutually independent homoskedastic errors independent of $\mathbf{U}$. Then all SEM parameters $\theta$ are identifiable from the joint distribution of $(\mathbf{X},\mathbf{Z},\mathbf{Y})$.
\end{Pro}

Reversing the signs of $\mathbf U$ and all three of its loadings preserves their products in Equation~\eqref{eq:sem1}. The condition $\beta_{UX}>0$ selects one of these two equivalent representations. To see the source of identification, write $a=\beta_{UX}$, $b=\beta_{UZ}$, and $d=\beta_{XY}a+\beta_{ZY}b+\beta_{UY}$. Then $\operatorname{Cov}(\mathbf{X})=a^2 \mathbf{C}+\sigma_X^2 \mathbf{I}$ and $\operatorname{Cov}(\mathbf{X},\mathbf{Y})=ad \mathbf{C}+\beta_{XY}\sigma_X^2 \mathbf{I}$. Linear independence of $\mathbf{C}$ and $\mathbf{I}$ uniquely separates their coefficients, identifying $a^2$, $\sigma_X^2$, $ad$, and $\beta_{XY}\sigma_X^2$. The remaining covariance blocks identify the other parameters, as shown in Appendix~\ref{app:proof}.

On an irregular tissue graph, the marginal variance of the field, $q_i=\mathbf C_{ii}$, can differ across receivers. Two distinct values of $q_i$ suffice for the linear independence in Proposition~\ref{prop1}, because a scalar multiple of $\mathbf I$ has a constant diagonal. Differences in marginal variance can therefore identify the effect through heteroskedasticity \citep{rigobon2003identification}. Even when all receivers have the same marginal variance, any nonzero covariance between distinct receivers also gives linear independence.

\paragraph{Priors.}
We assign a spike-and-slab prior to the ligand-exposure coefficient $\beta_{XY}$,
\begin{align*}
    \gamma &\sim \text{Bernoulli}(\pi_0), \\
    \beta_{XY} \mid \gamma &\sim \gamma \cdot \mathcal{N}(0, \tau^2) + (1-\gamma) \cdot \delta_0,
\end{align*}
where $\pi_0=0.5$, $\tau^2=1$, and $\delta_0$ is a point mass at zero. The posterior inclusion probability (PIP), $\Pr(\gamma=1\mid\mathrm{data})$, measures support for a nonzero exposure effect. Even a small effect can have a high PIP when it is precisely estimated. To assess magnitude, we report the model-averaged posterior mean and 95\% credible interval and the posterior probability that $|\beta_{XY}|$ exceeds 0.05, 0.10, or 0.20. An effect of 0.10 corresponds to a change of one-tenth of a pathway-score standard deviation per standard deviation of exposure. The remaining coefficients have independent normal priors,
\[
\beta_{ZY}, \beta_{UX}, \beta_{UZ}, \beta_{UY} \sim \mathcal{N}(0, \sigma_\beta^2),
\]
with $\sigma_\beta^2 = 1$ by default. The error variances have independent inverse-gamma priors,
\[
\sigma_X^2, \sigma_Z^2, \sigma_Y^2 \sim \text{Inverse-Gamma}(0.01, 0.01).
\]

\section{Posterior Computation and Diagnostics}\label{sec:computation}

We sample the coefficients, error variances, inclusion indicator, and latent field jointly using a blocked Gibbs sampler. Conditional on the coefficients and error variances, the three structural equations add only diagonal terms to the field precision matrix. The matrix therefore remains sparse, and hence we can sample the field for all receivers jointly by sparse Cholesky factorization. Appendix~\ref{app:posterior} gives the complete algorithm, including two Metropolis--Hastings moves used in the real-data fits.

The independent exposure variance $\sigma_X^2$ measures how much exposure variation remains after accounting for the shared field. A posterior concentrated near zero, or a strong posterior association between $\beta_{XY}$ and the outcome loading $\beta_{UY}$, can signal difficulty separating the exposure effect from the influence of the field. The sampler allows either sign for $\beta_{UX}$. To assess this association after sampling, we pair $\beta_{XY}$ in each retained draw with $\operatorname{sign}(\beta_{UX})\cdot\beta_{UY}$, the outcome loading under the sign convention of Proposition~\ref{prop1}. Sensitivity to the coefficient priors and neighborhood choices provides a further check. We assess the constant-variance sufficient condition by examining squared standardized residuals across receivers grouped by $q_i$ (Appendix~\ref{app:ovarian}).

\section{Real Data Analysis}\label{sec:real}
We compared CASPER and CellPhoneDB in four public 10x Genomics Xenium datasets, each representing one sample, from ovarian cancer \citep{tenx_xenium_ovarian}, melanoma \citep{tenx_xenium_skin_melanoma}, breast cancer \citep{tenx_xenium_breast}, and renal carcinoma \citep{tenx_xenium_renal}. We first asked how often the same candidates showed both ligand--receptor expression enrichment and evidence of an exposure effect on the receiver pathway. We then examined two examples in the ovarian tissue.

\subsection{Data Preprocessing and Screening}

We normalized raw counts to 10,000 per cell and applied the transformation $\log(1+x)$. Normalization used 5,101 genes in ovarian cancer and 405 in renal carcinoma. For melanoma and breast cancer it used 2,158 and 2,188 genes, respectively, drawn from Hallmark, ligand--receptor, and marker sets. We assigned broad cell types using dataset annotations and canonical markers. Each candidate specifies a ligand in the sender cell type, together with a receptor and a pathway in the receiver cell type. The ovarian examples use the Hallmark inflammatory-response and angiogenesis gene sets \citep{liberzon2015hallmark}. Some renal pathway gene sets were adapted to the smaller panel (Appendix~\ref{app:screening}).

Receptor expression $Z_i$ is the log-normalized receptor abundance. Pathway activity $Y_i$ is the mean log-normalized abundance of measured pathway genes, excluding the ligand and receptor. Before centering and scaling, ligand exposure is
\[
\widetilde X_i=\log\left\{1+\sum_{k\in\mathcal N_{32}(i)}
 \exp(-d_{ik}^2/2h_X^2)L_k\right\},
\]
where $L_k$ is normalized ligand abundance on the linear scale and $\mathcal N_{32}(i)$ comprises the 32 nearest sender cells. The bandwidth $h_X$ is the median positive distance among all retained sender--receiver neighbor distances for that candidate. %The receiver-cell GMRF uses $k=8$ and the analogous median-distance bandwidth $h_U$.

We considered pairs of distinct cell types with at least 100 cells of the sender type and 500 of the receiver type. We required log-normalized ligand and receptor expression above 0.1 in at least 5\% of their respective cell types, a product of their mean expressions of at least 0.01, and at least five measured downstream pathway genes after excluding ligand and receptor. In total, 2,229 candidates met these criteria. For ovarian, melanoma, and breast tissue, we ranked candidates using marginal correlation, spatial variation, and prior knowledge about the cell types and pathway. We used a specified list of eligible candidates for the renal panel. Appendix~\ref{app:screening} gives the selection rules.

Of the eligible candidates, 222 entered the comparison. We fitted CASPER using up to 3,000 receivers per candidate. For the two detailed ovarian analyses, we used all senders and receivers. The cancer-associated fibroblast VEGFA--FLT1 example belonged to the comparison set. The fibroblast IL6--IL6R example came from a separate exploratory search among other eligible ovarian candidates (Appendix~\ref{app:ovarian}).

We ran CellPhoneDB with 1,000 cell-label permutations, a 5\% expression threshold, and up to 3,000 cells per relevant type (Appendix~\ref{app:cellphonedb}). We used PIP $\geq0.7$ to identify candidates with high posterior support for a pathway effect and a nominal permutation $p<0.05$ to identify expression enrichment. A candidate could be untestable in CellPhoneDB if its interaction was missing from the database or a required gene product was absent from the input.

\subsection{Results}

\paragraph{Cross-dataset comparison.}
CellPhoneDB found expression enrichment for 151 of the 204 candidates it could test. CASPER gave high posterior support to a pathway effect for 14 of the 222 screened candidates (Table~\ref{tab:cpdb-comparison}). Nine of those 14 also showed expression enrichment. Of the remaining five, three involved WNT5A from keratinocyte/epithelial senders and one of three receptors (LRP6, FZD3, or FZD6) on melanoma mast cells. CellPhoneDB tested these three interactions and found no enrichment. The other two high-PIP candidates involved VEGFB--KDR in melanoma, for which the matched CellPhoneDB complex required FLT1, a component absent from the processed input.

\begin{table}[htbp]
\centering
\caption{Comparison of CASPER and CellPhoneDB by tissue. CASPER-high denotes PIP $\geq0.7$ for an effect on the receiver pathway. CellPhoneDB-positive denotes ligand--receptor expression enrichment at a nominal permutation $p<0.05$. The final column counts candidates meeting both criteria.}
\label{tab:cpdb-comparison}
\small
\begin{tabular}{lrrrrr}
\toprule
Dataset & Candidates & \shortstack[r]{CellPhoneDB\\tested} & \shortstack[r]{CASPER\\high} & \shortstack[r]{CellPhoneDB\\positive} & Both \\
\midrule
Breast & 20 & 20 & 2 & 13 & 2 \\
Melanoma & 89 & 79 & 6 & 55 & 1 \\
Ovarian & 92 & 92 & 3 & 70 & 3 \\
Renal & 21 & 13 & 3 & 13 & 3 \\
\bottomrule
\end{tabular}
\end{table}
\paragraph{Ovarian IL6--IL6R pathway response.}
We examined whether IL6 from 35,459 fibroblasts was associated with inflammation in 559 ciliated epithelial receivers. The score averaged 144 measured Hallmark inflammatory-response genes, and CASPER adjusted for receiver IL6R expression. The marginal correlation between exposure and pathway score was $-0.077$. CellPhoneDB found enrichment of fibroblast IL6 and the epithelial IL6R/IL6ST receptor complex, with no exceedances in 1,000 permutations.

CASPER assigned the IL6 effect a PIP of 0.924. A one-standard-deviation increase in exposure corresponded to an estimated increase of 0.708 standard deviations in pathway score under model averaging (95\% credible interval $[0,1.443]$). The interval includes zero because the posterior assigns 7.6\% probability to an exactly zero effect. Among draws with a nonzero effect, the 95\% interval was $[0.205,1.456]$. The posterior probability that the absolute effect exceeded 0.10 was 0.917 (Table~\ref{tab:supp-ovarian-effects}).

Expression of FOXJ1, CDHR4, and TEKT1 supported the ciliated epithelial annotation. IL6R has been localized to human fallopian-tube epithelial cilia \citep{shao2009cilia}, and IL6 exposure affects ciliary activity in human tubal explants \citep{papathanasiou2008cilia}. These observations make a response to IL6 plausible in this cell type. The score used here captures a broad inflammatory transcriptional response.

\paragraph{Ovarian VEGFA--FLT1 small-effect example.}
Our second example involved 33,990 cancer-associated fibroblasts and 2,794 angiogenic endothelial cells. We assessed exposure to fibroblast VEGFA and the activity of 21 measured angiogenesis genes in the receivers, adjusting for FLT1 expression. CellPhoneDB found enrichment for VEGFA--FLT1, with no exceedances in 1,000 permutations. The marginal exposure--pathway correlation was 0.010. CASPER assigned the effect a PIP of 0.053 and a model-averaged mean of $-0.0007$ (95\% credible interval $[-0.0181,0]$). The posterior probability of an absolute effect larger than 0.10 was 0.003. Thus the two pairs showed similar evidence of expression enrichment and markedly different evidence for a pathway response.

\paragraph{Diagnostics and sensitivity.}
Across the 36 receiver graphs used in the four-tissue comparison, the coefficient of variation of the marginal latent variances $q_i$ ranged from 0.56 to 4.65 (Table~\ref{tab:supp-q}). In the ovarian IL6 analysis, none of the three equations showed a clear linear trend in mean squared standardized residuals across ten groups of receivers ordered by $q_i$ (Appendix~\ref{app:ovarian}). When we shuffled IL6 exposure across receivers, the five control PIPs ranged from 0.043 to 0.283, and all five model-averaged intervals included zero.

The estimated IL6 effect was larger under a wider coefficient prior. The VEGFA estimate remained close to zero across the priors examined. Appendix~\ref{app:ovarian} reports the sensitivity analyses, and Table~\ref{tab:supp-ovarian-effects} summarizes the ovarian fits.

\section{Simulations}\label{sec:simulations}
We used simulations to examine how well CASPER estimates the ligand-exposure effect under different sources of spatial confounding. We compared CASPER with ordinary regression of $Y$ on $(X,Z)$, a spatial outcome-only regression with the same receiver graph and spike-and-slab prior, and an oracle regression that also observes the true confounder(s).

\subsection{Design}
We first examined how each method performed under the reference model as sample size and the true ligand-exposure effect varied. At each sample size ($n=500$, 3,000, or 10,000 receivers), we simulated a null effect ($\beta_{XY}=0$), a small effect (0.05), and a moderate effect (0.30). We generated the data from Equation~\eqref{eq:sem1} with all three confounder loadings set to 0.6, a receptor effect of 0.3, and independent errors of variance one. We sampled cell locations uniformly on a $10\times10$ domain and scaled the precision of the proper GMRF so that the mean marginal variance of the latent field was one.

We then fixed $n=500$ and examined departures from the reference model. We weakened or strengthened confounding by multiplying all three loadings by 0.5 or 1.5, respectively. We also examined two independent spatial fields with different ranges and nonproportional loadings, a quadratic term in the outcome equation, and independent variance-matched Laplace errors. Three additional settings assessed the effect of error variances that change with $q_i$. The exposure error variance increased in one setting and decreased in another. In the third setting, all three error variances increased. The ratio of the larger to the smaller variance was 1.5 across the interquartile range of $\log q_i$. Finally, we compared the irregular cell locations of the reference model with regular and jittered grids. We studied each of these variations at $\beta_{XY}=0$ and $0.30$. In each setting, we fitted all four methods to the same 100 independently generated datasets (Appendix~\ref{app:generators}).

CASPER and the spatial regression select an effect at PIP $\geq0.7$. Ordinary and oracle regressions select it when the 95\% confidence interval excludes zero. We evaluated how often each method selected the effect and how accurately it estimated its size. For bias and root mean squared error (RMSE), we transformed model-averaged Bayesian estimates back to the generating scale. For the posterior probabilities of exceeding an effect threshold, we used standardized coefficients.

\subsection{Results}
At $n=500$ under the reference model, CASPER selected a null effect in 3\% of replicates, compared with 42\% for spatial outcome-only regression, 93\% for ordinary regression, and 6\% for the oracle. For a true effect of 0.30, CASPER selected 97\% of replicates and had RMSE 0.071. The corresponding RMSEs were 0.142 for the spatial regression, 0.185 for ordinary regression, and 0.045 for the oracle. As the sample size increased to 3,000 and 10,000, CASPER detected the moderate effect in every replicate, and its RMSE fell to 0.022 and 0.013. The null selection rate was zero at both larger sample sizes.

The smaller effect of 0.05 illustrates how sample size can increase support for inclusion even when the estimated effect remains small. CASPER selected this effect in 3\%, 14\%, and 62\% of replicates at the three sample sizes. At $n=10{,}000$, the mean PIP was 0.727, while the mean posterior probability of an absolute standardized effect above 0.10 was 0.001.

At $n=500$, CASPER detected a true effect of 0.30 in 97--98\% of replicates under the two-field, mildly nonlinear, and Laplace scenarios, with absolute bias at most 0.018. Bias was greater when all three error variances increased with $q_i$. In that setting, CASPER underestimated the effect by 0.049 on average, its RMSE rose to 0.109, and it selected the effect in 85\% of replicates. Increasing the confounding multiplier from 0.5 to 1.5 also raised RMSE, from 0.056 to 0.098, and reduced selection from 100\% to 88\%.

\begin{figure}[htbp]
\centering
\includegraphics[width=\linewidth]{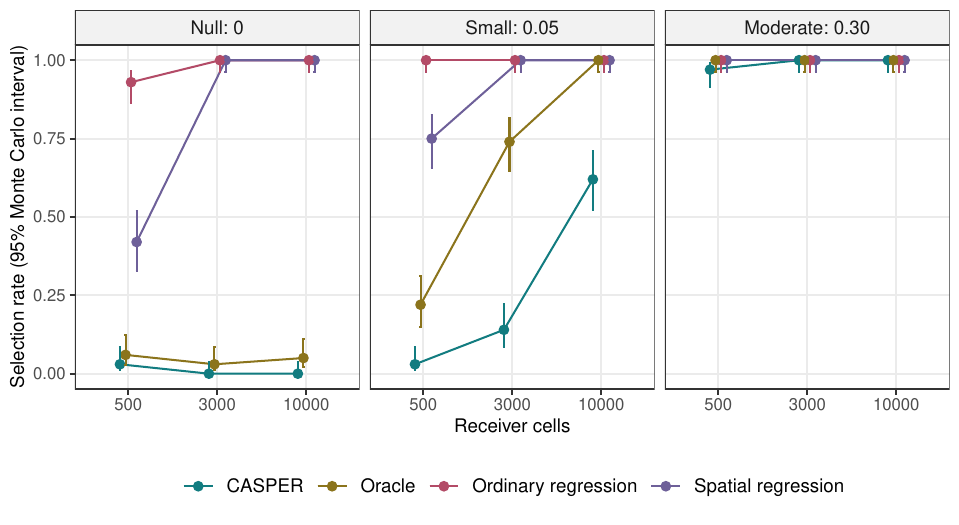}
\caption{Selection rates across 100 replications under the reference model, with 95\% Wilson intervals for Monte Carlo uncertainty. CASPER and spatial outcome-only regression select an effect when the PIP is at least 0.7. Ordinary and oracle regressions select it when the 95\% confidence interval excludes zero.}
\label{fig:sim-selection}
\end{figure}

On a regular periodic grid, the marginal latent variances were constant and the field still exhibited correlation between cells. CASPER selected the moderate effect in 91\% of replicates on this grid, in 88\% after jittering the locations, and in 97\% on irregular locations. The corresponding RMSEs were 0.090, 0.086, and 0.071.
Figure~\ref{fig:sim-selection} shows selection rates under the reference model. Table~\ref{tab:supp-simulations} gives the results across simulation settings.

\paragraph{Spatial-scale sensitivity.}
We refitted the same null and moderate-effect reference datasets at $n=500$ to examine sensitivity to the assumed spatial covariance. In separate fits we changed the number of receiver neighbors from eight to six or ten at the reference bandwidth, or we decreased or increased the bandwidth by 20\% at eight neighbors. Each covariance was scaled to unit mean marginal variance. Null selection remained between 2\% and 3\%. For the moderate effect, selection ranged from 90\% to 100\%, compared with 97\% at the reference settings, and the absolute shift in the mean effect estimate was at most 0.024 (Table~\ref{tab:supp-clean-scale}).

\section{Discussion}

CASPER estimates how a change in ligand exposure affects a receiver-cell pathway score at fixed receptor abundance. Modeling exposure, receptor expression, and pathway activity jointly allows the shared spatial field to explain variation in all three measurements. Under the linear model, independent homoskedastic errors and a known spatial covariance distinguishable from independent noise identify the effect.

The simulations illustrate both the value and the limits of spatial adjustment. Under the reference model, CASPER reduced the bias that persisted in ordinary and spatial outcome-only regressions. The bias remained small when the data were generated with two fields, a mild nonlinear effect of the confounder, or Laplace errors. Stronger confounding and error variances that increased with the marginal latent variance made estimation more difficult. The regular-grid experiment shows how correlation between cells can carry information even when their marginal variances are equal.

The ovarian examples illustrate the distinction between ligand--receptor expression enrichment and a response in the receiver pathway. CellPhoneDB found enrichment for both IL6 and VEGFA. CASPER gave considerably greater support to an effect of fibroblast IL6 on the ciliated epithelial inflammatory-response score than to an effect of fibroblast VEGFA on the endothelial angiogenesis score. 

The causal interpretation depends on how well a single spatial field captures shared neighborhood conditions within a receiver population. It also depends on the choice of exposure and pathway score. A negative coefficient means that greater exposure corresponds to a lower measured score after adjustment. The composition of the gene set or residual spatial confounding could contribute to such a result. Establishing that a particular receptor mediates a response would require additional biological evidence.

Future work could allow the ligand-exposure effect to vary across receiver subpopulations, with a hierarchical model borrowing information within a broader cell type. Joint models for exposures from several sender populations could help distinguish their contributions and examine competition or synergy through interaction terms. Propagating uncertainty in cell-type assignments through exposure construction and effect estimation would also address annotation error. These extensions raise further identification questions, particularly when sender exposures are strongly correlated or when effects vary across receivers.

\pdfbookmark[1]{References}{references}
\bibliographystyle{abbrvnat}
\bibliography{references}

\clearpage
\appendix
\setcounter{table}{0}
\renewcommand{\thetable}{S\arabic{table}}
\renewcommand{\theHtable}{supp.\arabic{table}}
\pdfbookmark[0]{Supplementary Material}{supplement}
\begin{center}
  {\Large\bfseries Supplementary Material}
\end{center}

\section{Proof of Proposition~\ref{prop1}}\label{app:proof}

The joint distribution of $(\mathbf X,\mathbf Z,\mathbf Y)$ determines all of their covariances. Whenever one of these covariances has the form $u\mathbf C+v\mathbf I$, it determines $u$ and $v$ separately because $\mathbf C$ and $\mathbf I$ are known and linearly independent. Consider first the exposure and receptor equations. By independence of the field and the errors,
\[
\operatorname{Var}(\mathbf X)=\beta_{UX}^2\mathbf C+\sigma_X^2\mathbf I,
\qquad
\operatorname{Var}(\mathbf Z)=\beta_{UZ}^2\mathbf C+\sigma_Z^2\mathbf I.
\]
The coefficients of $\mathbf I$ give the two error variances. By the sign assumption, $\beta_{UX}$ is the positive square root of the coefficient of $\mathbf C$ in $\operatorname{Var}(\mathbf X)$. The cross-covariance $\operatorname{Cov}(\mathbf X,\mathbf Z)=\beta_{UX}\beta_{UZ}\mathbf C$ then determines $\beta_{UZ}$, including its sign.

To recover the two direct effects on pathway activity, consider the residual $\mathbf R(b,c)=\mathbf Y-b\mathbf X-c\mathbf Z$. At $b=\beta_{XY}$ and $c=\beta_{ZY}$, the outcome equation gives $\mathbf R(b,c)=\beta_{UY}\mathbf U+\boldsymbol\epsilon_Y$. Its covariances with exposure and receptor expression are therefore both proportional to $\mathbf C$. Conversely, suppose both cross-covariances are proportional to $\mathbf C$ for some $b$ and $c$. The outcome equation gives
\[
\mathbf R(b,c)=(\beta_{XY}-b)\mathbf X+(\beta_{ZY}-c)\mathbf Z
                +\beta_{UY}\mathbf U+\boldsymbol\epsilon_Y.
\]
In the cross-covariances of this residual with $\mathbf X$ and $\mathbf Z$, the coefficients of $\mathbf I$ are $(\beta_{XY}-b)\sigma_X^2$ and $(\beta_{ZY}-c)\sigma_Z^2$, respectively. Both coefficients must vanish. Since the error variances are positive, this forces $b=\beta_{XY}$ and $c=\beta_{ZY}$. Thus the joint law determines the two direct effects.

For the residual $\mathbf R=\mathbf R(\beta_{XY},\beta_{ZY})$, independence of the outcome error gives
\[
\operatorname{Cov}(\mathbf X,\mathbf R)=\beta_{UX}\beta_{UY}\mathbf C,
\qquad
\operatorname{Var}(\mathbf R)=\beta_{UY}^2\mathbf C+\sigma_Y^2\mathbf I.
\]
The first identity gives $\beta_{UY}$ because $\beta_{UX}$ is known and positive. The only unknown in the second identity is then $\sigma_Y^2$, which completes the proof. \qed

Variation in the diagonal entries $q_i=\mathbf C_{ii}$ is sufficient for linear independence. With a constant diagonal, any nonzero off-diagonal covariance also gives linear independence. For the marginal variance $\operatorname{Var}(X_i)=\beta_{UX}^2q_i+\sigma_X^2$, a narrow range of $q_i$ makes the slope harder to distinguish from the intercept. The full likelihood also uses the off-diagonal entries of $\mathbf C$. Appendix~\ref{app:generators} examines estimation on graphs with varying degrees of regularity.

\section{Posterior Computation}\label{app:posterior}
The sampler updates the shared spatial field, the regression coefficients, the inclusion indicator for the exposure effect, and the three error variances. The conditional distribution of the spatial field is Gaussian. We use sparse Cholesky factorization to draw from this distribution under the receiver-graph precision in Section~\ref{sec:method} or the simulation precisions in Appendix~\ref{app:generators}.

\paragraph{Blocked outcome update.}
Conditional on $\mathbf{U}$ and $\sigma_Y^2$, let $\mathbf{W}_0=[\mathbf{Z},\mathbf{U}]$ and $\mathbf{W}_1=[\mathbf{X},\mathbf{Z},\mathbf{U}]$, with $p_0=2$ and $p_1=3$. Included outcome coefficients have independent $N(0,\tau^2)$ priors with $\tau^2=1$ by default. Appendix~\ref{app:ovarian} examines how the ovarian estimates change with this common variance. For $g\in\{0,1\}$ define
\[
\mathbf{V}_g=(\mathbf{W}_g^\top \mathbf{W}_g/\sigma_Y^2+\tau^{-2}\mathbf{I})^{-1},
\qquad \mathbf{h}_g=\mathbf{W}_g^\top \mathbf{Y}/\sigma_Y^2.
\]
Apart from constants common to both models, the integrated log likelihood is
\[
\ell_g=\tfrac12\{\log|\mathbf{V}_g|-p_g\log(\tau^2)+\mathbf{h}_g^\top \mathbf{V}_g\mathbf{h}_g\}.
\]
With prior inclusion probability 0.5, the posterior conditional probability of $\gamma=1$ is $\{1+\exp(\ell_0-\ell_1)\}^{-1}$. Given the sampled value of $\gamma$, the included coefficients have distribution $N(\mathbf{V}_\gamma\mathbf{h}_\gamma,\mathbf{V}_\gamma)$. When $\gamma=0$, $\beta_{XY}=0$.

\paragraph{Remaining conjugate updates.}
For $A\in\{X,Z\}$, the loading has a normal full conditional with variance $V_A=(\mathbf U^\top\mathbf U/\sigma_A^2+\sigma_\beta^{-2})^{-1}$ and mean $V_A\mathbf U^\top\mathbf A/\sigma_A^2$, where $\mathbf A$ denotes the corresponding observed vector and $\sigma_\beta^2=1$ by default. Conditional on the other variables,
\[
\sigma_A^2\mid\mathrm{rest}\sim\operatorname{Inverse\text{-}Gamma}
\left(0.01+\frac n2,\;0.01+\frac12\|\mathbf r_A\|^2\right),\qquad A\in\{X,Z,Y\},
\]
where $\mathbf r_X=\mathbf X-\beta_{UX}\mathbf U$, $\mathbf r_Z=\mathbf Z-\beta_{UZ}\mathbf U$, and $\mathbf r_Y=\mathbf Y-\beta_{XY}\mathbf X-\beta_{ZY}\mathbf Z-\beta_{UY}\mathbf U$. In the real-data analyses, we update the exposure variance after integrating out the field, as detailed below.

\paragraph{Additional updates for the real-data analyses.}
Two updates improve mixing when the data provide limited information about the scale of the field and its loadings. The first samples the exposure variance $v=\sigma_X^2$ after integrating out $\mathbf U$. Write $\mathbf y_*=\mathbf Y-\beta_{XY}\mathbf X-\beta_{ZY}\mathbf Z$. Given the data and the equation parameters, the conditional distribution of the field has precision $\mathbf A(v)$ and mean $\mathbf m(v)$, where
\[
\begin{aligned}
\mathbf A(v)&=\mathbf Q+(\beta_{UX}^2/v+\beta_{UZ}^2/\sigma_Z^2+\beta_{UY}^2/\sigma_Y^2)\mathbf I,\\
\mathbf m(v)&=\mathbf A(v)^{-1}(\beta_{UX}\mathbf X/v+\beta_{UZ}\mathbf Z/\sigma_Z^2+\beta_{UY}\mathbf y_*/\sigma_Y^2).
\end{aligned}
\]
For $t=\log v$, the marginal log conditional density, up to an additive constant, is
\[
-\tfrac12\{nt+\log|\mathbf A(e^t)|+J(e^t)\}-0.01t-0.01e^{-t},
\]
where, writing $\mathbf m=\mathbf m(v)$,
\[
J(v)=\|\mathbf X-\beta_{UX}\mathbf m\|^2/v+
\|\mathbf Z-\beta_{UZ}\mathbf m\|^2/\sigma_Z^2+
\|\mathbf y_*-\beta_{UY}\mathbf m\|^2/\sigma_Y^2+\mathbf m^\top\mathbf Q\mathbf m.
\]
A Metropolis--Hastings step uses a symmetric normal random-walk proposal for $t$, with the proposal scale adapted during warm-up. We then draw $\mathbf U$ from its conditional distribution at the resulting value of $v$.

The second update rescales the field and all three loadings together. For a proposed log-scale change $s\sim N(0,1.5^2/n)$, the field is multiplied by $e^s$ and each loading by $e^{-s}$. These changes preserve their products in the likelihood. Writing $\boldsymbol\beta_U$ for the vector of three loadings, the log acceptance ratio is
\[
-\tfrac12(e^{2s}-1)\mathbf U^\top\mathbf Q\mathbf U
-\frac{1}{2\sigma_\beta^2}(e^{-2s}-1)\|\boldsymbol\beta_U\|^2+(n-3)s,
\]
where the last term is the log Jacobian.

\paragraph{Posterior summaries.}
Simulation fits used one chain with 1,000 warm-up iterations and 2,000 retained draws, saving every second draw. The four-tissue comparison used one chain per candidate with 1,500 warm-up iterations and 2,000 retained draws without thinning. Each detailed ovarian fit used four independent chains and 3,000 warm-up iterations per chain. We retained 16,000 subsequent draws per chain for IL6--IL6R and 4,000 for VEGFA--FLT1.

The posterior inclusion probability (PIP) is the fraction of draws with $\gamma=1$. Model-averaged 95\% credible intervals use the empirical 2.5th and 97.5th percentiles of all draws, including the zeros. Conditional intervals use only draws with $\gamma=1$. At each threshold $\delta$, we estimate $\Pr(|\beta_{XY}|>\delta\mid\mathrm{data})$ by the fraction of all draws whose absolute value exceeds $\delta$. Before computing convergence diagnostics for the field loadings, we multiply all three loadings in each retained draw by $\operatorname{sign}(\beta_{UX})$.

\section{Real-Data Methods}\label{app:data}

\subsection{Candidate Screening}\label{app:screening}
We first identified ligand--receptor pairs for which both genes and a downstream pathway could be measured in the relevant cell types. A candidate linked two distinct cell types with at least 100 sender cells and 500 receiver cells. Log-normalized abundance had to exceed 0.1 in at least 5\% of the respective cells for both the ligand and the receptor. The product $S_{\mathrm{expr}}$ of mean sender ligand abundance and mean receiver receptor abundance had to reach 0.01, and at least five measured pathway genes had to remain after excluding the ligand and receptor.

Of the 2,229 eligible candidates, we ranked the 300 ovarian candidates with the largest $S_{\mathrm{expr}}$ (breaking ties by pathway size) and all eligible melanoma and breast candidates. We calculated the screening scores from the transformed measurements before standardizing them for model fitting. Four rankings allowed us to include candidates with strong marginal association, substantial spatial variation, prior biological support, or a spatial pattern despite weak marginal association. Write $r$ for the marginal exposure--pathway correlation, $G$ for the number of measured pathway genes, and $I_X,I_Z,I_Y$ for Moran's $I$ of exposure, receptor abundance, and pathway activity among receivers. With $u_+=\max(u,0)$, the rankings use
\[
\begin{aligned}
A&=\log(1+S_{\mathrm{expr}})\log(1+G),\qquad
B=\log\{1+\operatorname{Var}(X)+\operatorname{Var}(Z)+\operatorname{Var}(Y)\},\\
C&=1+(I_X)_++(I_Z)_++(I_Y)_+,\qquad S=AC(1+B).
\end{aligned}
\]
The four scores, in order, were $|r|AC$, $S$, $S w_{\mathrm{bio}}$, and $S/(1+10|r|)$. The screening code specifies the biological weights $w_{\mathrm{bio}}$ by pathway, genes, and sender--receiver pairing. The fourth ranking was restricted to $|r|\leq0.10$. The union of the top 50 candidates per ranking in ovarian tissue and melanoma and the top 10 in breast tissue gave 92, 89, and 20 distinct candidates, respectively. A specified list of 21 eligible renal candidates brought the total to 222. The analysis code provides the genes defining each pathway outcome, including panel-adapted gene sets for the renal tissue.

Table~\ref{tab:supp-q} summarizes the quartiles and the coefficient of variation of the marginal latent variances $q_i=(\mathbf Q^{-1})_{ii}$ for each of the 36 receiver graphs used in the comparison. The coefficients of variation ranged from 0.56 to 4.65.

\subsection{CellPhoneDB Comparison}\label{app:cellphonedb}
CellPhoneDB compares expression of a ligand--receptor interaction in a cell-type pair with its expression under permuted cell-type labels \citep{efremova2020cellphonedb}. We used version 5.0.1 with database v5.0.0, 1,000 label permutations, and a 5\% expression prevalence threshold. Each tissue analysis used all retained RNA genes and up to 3,000 cells per relevant annotated type.

We matched ligand and receptor genes to opposite sides of a database interaction in either orientation. Matches could involve an exact pair or a complex containing a candidate gene. A complex could be tested only if all required components were present. Among multiple matches, we gave priority to tested interactions, then to exact pairs, breaking ties by database identifier. We recorded missing components and absent database matches separately. Candidates with the same ligand, receptor, and cell types share a CellPhoneDB result across pathways. We used a nominal permutation $p<0.05$ for enrichment. The comparison describes candidates selected using expression and spatial information from the same tissues.

\subsection{Ovarian Follow-up}\label{app:ovarian}
The two ovarian examples had different origins. VEGFA--FLT1 was part of the four-tissue comparison. IL6--IL6R arose in a separate exploratory search among expression-eligible ovarian cytokines. For that search, we profiled the exposure covariance $\operatorname{Cov}(\mathbf X)=\beta_{UX}^2\mathbf Q^{-1}+\sigma_X^2\mathbf I$. We fitted the joint model for candidates whose profile had an interior maximum and whose estimated independent exposure variance was at least 0.05 on the standardized scale.

Both follow-ups used all cells of the designated sender and receiver types. The sender-exposure and receiver-graph bandwidths were 111 and 134 micrometers, respectively, for IL6 and 49 and 47 micrometers for VEGFA. We standardized exposure, receptor abundance, and pathway score within each receiver population. Only 94 of 559 ciliated epithelial receivers had positive fibroblast IL6 exposure before centering. Across all 559 IL6 receivers, the median $q_i$ was 2.60 (interquartile range 2.06--4.80), compared with 4.17 (2.58--8.93) for the VEGFA receivers. Table~\ref{tab:supp-ovarian-effects} gives the posterior summaries. Across the two fits, the largest rank-normalized split $\widehat R$ for continuous parameters was 1.008. The smallest bulk and tail effective sample sizes were 473 and 1,047. The Monte Carlo standard errors of the PIPs were 0.0039 for IL6 and 0.0019 for VEGFA.

\paragraph{Residual variance in the IL6 analysis.}
We examined whether residual variance changed with $q_i$ in any of the three equations. At each of the 64,000 retained draws, let $\mu_i$ and $V_{ii}$ be the conditional mean and variance of $U_i$ given the data and parameters, as in Appendix~\ref{app:posterior}. Write $t_{Xi}=X_i$, $t_{Zi}=Z_i$, and $t_{Yi}=Y_i-\beta_{XY}X_i-\beta_{ZY}Z_i$. The conditional expected squared standardized residual for equation $A\in\{X,Z,Y\}$ was
\[
\frac{(t_{Ai}-\beta_{UA}\mu_i)^2+\beta_{UA}^2 V_{ii}}{\sigma_A^2}.
\]
We averaged this quantity over draws for each receiver, then within ten groups of nearly equal size ordered by $q_i$ (55 or 56 cells, with ties resolved by receiver order). For each equation we regressed the ten group averages on group mean $q_i$ with an intercept. Per receiver-level standard deviation of $q_i$ (9.05), the estimated slopes were $-0.029$ for IL6 exposure ($p=0.808$), $0.154$ for IL6R expression ($p=0.252$), and $0.003$ for the inflammatory-response score ($p=0.978$). 

\paragraph{Exposure-permutation controls.}
We randomly reassigned the standardized IL6 exposures across the 559 receivers five times, holding receptor abundance, pathway scores, coordinates, and the receiver graph fixed. Each fit used the original priors and chain lengths. All five model-averaged 95\% intervals included zero, and the PIPs ranged from 0.043 to 0.283. These controls assess how the result for this candidate depends on the spatial alignment of exposure with receiver cells.

\paragraph{Sensitivity to coefficient priors.}
We refitted both examples with common normal coefficient variance 0.25 or 4 in place of 1, including the slab variance for $\beta_{XY}$. Each alternative used four chains with 16,000 retained draws per chain. As the variance increased from 0.25 to 4, the IL6 PIP rose from 0.823 to 0.951 and the posterior mean conditional on inclusion rose from 0.453 to 1.136. Across all three priors the conditional intervals were positive and the model-averaged intervals included zero. For VEGFA the PIP ranged from 0.027 to 0.098 and the conditional intervals remained close to $[-0.12,0.09]$.

{\footnotesize
\begin{longtable}{@{}llrr@{}}
\caption{Distribution of the marginal latent variance $q_i$ within each of the 36 receiver graphs. Each graph contains up to 3,000 cells. Q1 and Q3 denote the 25th and 75th percentiles. The coefficient of variation (CV) is the standard deviation divided by the mean.}\label{tab:supp-q}\\
\toprule
Tissue & Receiver cell type & Median [Q1, Q3] & CV \\
\midrule
\endfirsthead
\multicolumn{4}{@{}l}{\emph{Table~\thetable{} (continued)}}\\
\toprule
Tissue & Receiver cell type & Median [Q1, Q3] & CV \\
\midrule
\endhead
Ovarian & Angiogenic endothelial & 4.17 [2.58, 8.93] & 1.59 \\
 & CAF & 0.93 [0.69, 1.34] & 3.61 \\
 & Cycling epithelial or unknown & 0.95 [0.73, 1.35] & 2.99 \\
 & Cycling tumor epithelial & 0.87 [0.72, 1.21] & 3.02 \\
 & Endothelial & 0.88 [0.71, 1.18] & 3.23 \\
 & Fibroblast & 0.76 [0.52, 1.30] & 3.40 \\
 & Hypoxic tumor or stress & 1.43 [1.06, 2.23] & 2.92 \\
 & Inflammatory fibroblast & 1.13 [0.91, 1.70] & 3.27 \\
 & Macrophage & 0.90 [0.68, 1.49] & 1.77 \\
 & Mesothelial & 2.13 [1.49, 4.07] & 2.43 \\
 & Myeloid & 0.80 [0.59, 1.16] & 4.19 \\
 & Pericyte & 3.72 [2.17, 7.32] & 1.82 \\
 & Tumor epithelial & 0.96 [0.79, 1.29] & 3.40 \\
\addlinespace[2pt]
Melanoma & B cell & 1.06 [0.92, 1.50] & 1.59 \\
 & Endothelial & 1.59 [1.00, 2.81] & 1.89 \\
 & Fibroblast CAF & 0.68 [0.56, 1.06] & 3.01 \\
 & Keratinocyte epithelial & 0.60 [0.55, 0.73] & 3.25 \\
 & Mast cell & 0.75 [0.60, 1.24] & 2.83 \\
 & Melanoma melanocytic & 0.65 [0.55, 0.83] & 4.65 \\
 & Myeloid & 0.79 [0.61, 1.07] & 2.92 \\
 & Perivascular smooth muscle & 0.62 [0.54, 0.87] & 2.61 \\
 & Plasma cell & 0.89 [0.66, 2.05] & 2.34 \\
 & Proliferating melanoma & 0.71 [0.60, 0.90] & 4.33 \\
 & T NK cell & 0.88 [0.59, 1.77] & 2.28 \\
\addlinespace[2pt]
Renal & B plasma & 1.37 [0.90, 4.13] & 2.42 \\
 & B plasma mixed & 1.20 [0.86, 2.12] & 2.74 \\
 & Endothelial & 0.69 [0.60, 0.82] & 0.56 \\
 & Stromal & 0.67 [0.53, 1.06] & 1.39 \\
 & Stromal mixed & 0.68 [0.51, 1.00] & 0.83 \\
 & T cell & 0.91 [0.70, 1.20] & 0.91 \\
 & T cell mixed & 0.73 [0.62, 0.89] & 1.15 \\
\addlinespace[2pt]
Breast & Endothelial & 0.68 [0.56, 0.90] & 2.68 \\
 & Mast cell & 0.76 [0.61, 1.03] & 3.59 \\
 & Perivascular & 0.85 [0.71, 1.24] & 2.37 \\
 & Plasma cell & 0.92 [0.68, 1.36] & 3.35 \\
 & Tumor epithelial & 1.17 [0.79, 1.96] & 2.60 \\
\bottomrule
\end{longtable}
}
\begin{table}[H]
\centering\small
\caption{Posterior summaries for the ovarian examples. Effects are in pathway-score standard deviations per exposure standard deviation. Model-averaged means and 95\% credible intervals use all draws. Conditional intervals use draws with $\gamma=1$. $P_\delta$ is the posterior probability that $|\beta_{XY}|>\delta$.}
\label{tab:supp-ovarian-effects}
\begin{tabular}{@{}lrr@{}}\toprule
Posterior summary & IL6--IL6R & VEGFA--FLT1 \\ \midrule
Inclusion probability & 0.924 & 0.053 \\
Model-averaged mean & 0.708 & $-0.0007$ \\
Model-averaged 95\% CrI & $[0,1.443]$ & $[-0.018,0]$ \\
95\% CrI given inclusion & $[0.205,1.456]$ & $[-0.116,0.087]$ \\ \addlinespace
$P_{0.05}$ & 0.921 & 0.019 \\
$P_{0.10}$ & 0.917 & 0.003 \\
$P_{0.20}$ & 0.903 & 0.000 \\ \bottomrule
\end{tabular}
\end{table}

\section{Simulation Generators}\label{app:generators}
\subsection{Simulation Design}\label{app:design}
The reference experiment asks how well the methods recover a known exposure effect when exposure, receptor expression, and pathway activity share a spatial cause. The remaining experiments change the strength or form of that confounding, the error variances, or the geometry of the tissue. Under the reference model, $X=0.6U+\epsilon_X$, $Z=0.6U+\epsilon_Z$, and $Y=\beta X+0.3Z+0.6U+\epsilon_Y$, with independent unit-variance Gaussian errors. The weak- and strong-confounding settings multiply the three latent loadings by 0.5 and 1.5, respectively. Table~\ref{tab:supp-simulations} summarizes the simulations. 

The reference locations were sampled uniformly from a square. To set the scale of the Gaussian field, we started with the receiver-graph precision $\mathbf Q_0=\mathbf D-\mathbf W+0.01\mathbf I$, where $\mathbf W$ is the weighted adjacency matrix and $\mathbf D$ the diagonal matrix of weighted degrees. The scaled precision is $\mathbf Q=c\mathbf Q_0$, with $c=n^{-1}\operatorname{tr}(\mathbf Q_0^{-1})$. This gives mean marginal variance one. We calculated $c$ from the exact diagonal for $n\leq3{,}000$ and estimated it at $n=10{,}000$ using the mean squared norm of 64 draws from $N(\mathbf 0,\mathbf Q_0^{-1})$, divided by $n$. Below, $q=(q_1,\ldots,q_n)^\top$ collects the marginal variances $q_i=(\mathbf Q^{-1})_{ii}$.

In the two-field experiment, independent $U_1$ and $U_2$ have unit average marginal variance. The first graph uses the reference Gaussian bandwidth $h_U$, the median positive distance among receiver-neighbor pairs, and the second uses $2h_U$. Their contributions to $X$, $Z$, and $Y$ are respectively
\[
0.6(\sqrt{0.75}U_1+\sqrt{0.25}U_2),\quad
0.6(\sqrt{0.25}U_1+\sqrt{0.75}U_2),\quad
0.6(\sqrt{0.5}U_1+\sqrt{0.5}U_2).
\]
CASPER fits a single field using the first precision. The oracle observes both fields. The mild nonlinear experiment changes only the direct confounding contribution to $Y$, to
\[
0.6\{U+0.1(U^2-q)/\sqrt{2\overline{q^2}}\}/\sqrt{1.01},
\]
where the square is elementwise and $\overline{q^2}=n^{-1}\sum_i q_i^2$. The oracle includes both $U$ and the quadratic term. The Laplace experiment replaces each error by an independent random sign times an exponential variable with rate $\sqrt2$, preserving variance one.

For heteroskedasticity, define
\[
t_i=\frac{\log q_i-\operatorname{median}(\log q)}{\operatorname{IQR}(\log q)},
\qquad v_i=\exp\{\pm\log(1.5)t_i\},
\]
and normalize the resulting $v_i$ to have mean one. Multiplying the corresponding errors by $\sqrt{v_i}$ scales their variances by $v_i$. In two settings, the variance in the exposure equation increased or decreased with $q_i$. In a third, the error variance in all three equations increased with $q_i$. CASPER assumed constant error variances in each fit.

The regular graph has 500 cells on a periodic $20\times25$ lattice, with eight neighbors and symmetrized Gaussian weights. Its marginal variances are constant to numerical precision. Jitter adds independent uniform coordinate offsets up to 20\% of the corresponding lattice spacing and wraps coordinates periodically. The irregular comparator is the reference random point pattern. The periodic graph retains off-diagonal spatial covariance even when the diagonal is constant. Regression estimates include an intercept. Bayesian fits use centered, standardized variables, and we back-transform effect estimates to the generating scale to calculate bias and root mean squared error.

\subsection{Sensitivity to the fitted spatial covariance}\label{app:sensitivity}
We varied the receiver graph used to fit CASPER while holding each dataset, including its exposure, fixed. For each generating effect ($\beta=0$ or $0.30$), we refitted 100 reference datasets at $n=500$. At the reference bandwidth $h_0$ we changed the number of neighbors from eight to six or ten. With eight neighbors we used $0.8h_0$ or $1.2h_0$. Each fitted covariance had mean marginal variance one. Table~\ref{tab:supp-clean-scale} shows how these changes affected selection and the estimated exposure effect.

{\footnotesize
\begin{longtable}{@{}>{\raggedright\arraybackslash}p{3.6cm}rlrrrr@{}}
\caption{Selection and estimation in 100 replicates per setting. CASPER and spatial outcome-only regression select effects at posterior inclusion probability (PIP) $\geq 0.7$. Ordinary and oracle regression use a 95\% confidence interval excluding zero. Bias and root mean squared error (RMSE) are on the generating scale.}\label{tab:supp-simulations}\\
\toprule
Setting & Effect & Method & Selection & Mean PIP & Bias & RMSE \\
\midrule
\endfirsthead
\multicolumn{7}{@{}l}{\emph{Table~\thetable{} (continued)}}\\
\toprule
Setting & Effect & Method & Selection & Mean PIP & Bias & RMSE \\
\midrule
\endhead
\multicolumn{7}{@{}l}{\emph{Reference model}} \\*
$n=500$ & 0.00 & CASPER & 0.03 & 0.101 & 0.001 & 0.027 \\*
 &  & Spatial & 0.42 & 0.540 & 0.078 & 0.103 \\*
 &  & Ordinary & 0.93 & --- & 0.169 & 0.177 \\*
 &  & Oracle & 0.06 & --- & -0.003 & 0.048 \\[2pt]
 & 0.05 & CASPER & 0.03 & 0.147 & -0.036 & 0.047 \\*
 &  & Spatial & 0.75 & 0.822 & 0.096 & 0.118 \\*
 &  & Ordinary & 1.00 & --- & 0.161 & 0.171 \\*
 &  & Oracle & 0.22 & --- & -0.004 & 0.045 \\[2pt]
 & 0.30 & CASPER & 0.97 & 0.968 & -0.001 & 0.071 \\*
 &  & Spatial & 1.00 & 1.000 & 0.134 & 0.142 \\*
 &  & Ordinary & 1.00 & --- & 0.179 & 0.185 \\*
 &  & Oracle & 1.00 & --- & 0.009 & 0.045 \\[2pt]
$n=3{,}000$ & 0.00 & CASPER & 0.00 & 0.041 & 0.000 & 0.004 \\*
 &  & Spatial & 1.00 & 1.000 & 0.137 & 0.138 \\*
 &  & Ordinary & 1.00 & --- & 0.198 & 0.199 \\*
 &  & Oracle & 0.03 & --- & 0.000 & 0.016 \\[2pt]
 & 0.05 & CASPER & 0.14 & 0.284 & -0.029 & 0.041 \\*
 &  & Spatial & 1.00 & 1.000 & 0.136 & 0.138 \\*
 &  & Ordinary & 1.00 & --- & 0.201 & 0.202 \\*
 &  & Oracle & 0.74 & --- & -0.002 & 0.020 \\[2pt]
 & 0.30 & CASPER & 1.00 & 1.000 & -0.001 & 0.022 \\*
 &  & Spatial & 1.00 & 1.000 & 0.137 & 0.138 \\*
 &  & Ordinary & 1.00 & --- & 0.200 & 0.201 \\*
 &  & Oracle & 1.00 & --- & -0.001 & 0.018 \\[2pt]
$n=10{,}000$ & 0.00 & CASPER & 0.00 & 0.028 & 0.000 & 0.001 \\*
 &  & Spatial & 1.00 & 1.000 & 0.144 & 0.144 \\*
 &  & Ordinary & 1.00 & --- & 0.209 & 0.209 \\*
 &  & Oracle & 0.05 & --- & 0.001 & 0.010 \\[2pt]
 & 0.05 & CASPER & 0.62 & 0.727 & -0.010 & 0.026 \\*
 &  & Spatial & 1.00 & 1.000 & 0.142 & 0.142 \\*
 &  & Ordinary & 1.00 & --- & 0.207 & 0.207 \\*
 &  & Oracle & 1.00 & --- & 0.000 & 0.010 \\[2pt]
 & 0.30 & CASPER & 1.00 & 1.000 & 0.002 & 0.013 \\*
 &  & Spatial & 1.00 & 1.000 & 0.144 & 0.144 \\*
 &  & Ordinary & 1.00 & --- & 0.208 & 0.208 \\*
 &  & Oracle & 1.00 & --- & 0.001 & 0.010 \\[2pt]
\midrule
\multicolumn{7}{@{}l}{\emph{CASPER under perturbations at $n=500$}} \\*
Weaker confounding & 0.00 & CASPER & 0.01 & 0.087 & -0.001 & 0.015 \\*
 & 0.30 &  & 1.00 & 0.993 & -0.004 & 0.056 \\[2pt]
Stronger confounding & 0.00 &  & 0.00 & 0.087 & 0.000 & 0.010 \\*
 & 0.30 &  & 0.88 & 0.906 & -0.021 & 0.098 \\[2pt]
Two spatial fields & 0.00 &  & 0.00 & 0.085 & 0.002 & 0.011 \\*
 & 0.30 &  & 0.98 & 0.983 & 0.001 & 0.057 \\[2pt]
Nonlinear outcome & 0.00 &  & 0.00 & 0.082 & -0.001 & 0.015 \\*
 & 0.30 &  & 0.97 & 0.971 & -0.002 & 0.068 \\[2pt]
Laplace errors & 0.00 &  & 0.01 & 0.111 & 0.002 & 0.020 \\*
 & 0.30 &  & 0.97 & 0.967 & -0.017 & 0.066 \\[2pt]
Increasing $X$ error variance & 0.00 &  & 0.00 & 0.092 & -0.002 & 0.015 \\*
 & 0.30 &  & 0.92 & 0.939 & -0.027 & 0.082 \\[2pt]
Decreasing $X$ error variance & 0.00 &  & 0.00 & 0.076 & 0.001 & 0.006 \\*
 & 0.30 &  & 0.97 & 0.979 & 0.006 & 0.066 \\[2pt]
Increasing error variance in all three equations & 0.00 &  & 0.04 & 0.134 & -0.004 & 0.040 \\*
 & 0.30 &  & 0.85 & 0.886 & -0.049 & 0.109 \\[2pt]
Regular grid & 0.00 &  & 0.00 & 0.122 & -0.001 & 0.028 \\*
 & 0.30 &  & 0.91 & 0.911 & -0.020 & 0.090 \\[2pt]
Jittered grid & 0.00 &  & 0.02 & 0.118 & 0.005 & 0.031 \\*
 & 0.30 &  & 0.88 & 0.911 & -0.022 & 0.086 \\[2pt]
\bottomrule
\end{longtable}
}
\begin{table}[H]
\centering\small
\caption{Sensitivity to the fitted receiver graph in 100 datasets per generating effect. Selection is the proportion of fits with PIP $\geq0.7$. At eight neighbors and bandwidth $h_0$, the selection rates were 0.03 ($\beta=0$) and 0.97 ($\beta=0.30$). For each alternative, $\Delta$ is the average within-dataset change in the model-averaged exposure effect from this reference fit, on the generating scale. Parentheses give its Monte Carlo standard error.}
\label{tab:supp-clean-scale}
\begin{tabular}{@{}lrrrr@{}}\toprule
 & \multicolumn{2}{c}{$\beta=0$} & \multicolumn{2}{c}{$\beta=0.30$} \\
\cmidrule(lr){2-3}\cmidrule(l){4-5}
Fitted graph & Selection & $\Delta$ (MCSE) & Selection & $\Delta$ (MCSE) \\ \midrule
$k=6$ & 0.03 & 0.003 (0.001) & 1.00 & 0.023 (0.001) \\
$k=10$ & 0.02 & -0.003 (0.001) & 0.90 & -0.018 (0.002) \\
$h=0.8h_0$ & 0.03 & 0.002 (0.000) & 1.00 & 0.017 (0.001) \\
$h=1.2h_0$ & 0.03 & -0.001 (0.000) & 0.93 & -0.011 (0.001) \\
\bottomrule
\end{tabular}
\end{table}

\end{document}